\documentclass[10pt,aps,prd,twocolumn]{revtex4-1}
\usepackage{acronym,amsmath,graphicx,textcomp,hyperref}

\begin{document}
 
\title{First searches for gravitational waves from r-modes of the Crab pulsar}
\author{Binod Rajbhandari}
\affiliation{
Department of Physics and Astronomy,
Texas Tech University,
Lubbock, Texas, 79409-1051, USA
} 
\author{Benjamin J. Owen}
\affiliation{
Department of Physics and Astronomy,
Texas Tech University,
Lubbock, Texas, 79409-1051, USA
}
\author{Santiago Caride}
\altaffiliation[Current address:]{
Target Field,
1 Twins Way,
Minneapolis,MN, 55403
}
\affiliation{
Department of Physics and Astronomy,
Texas Tech University,
Lubbock, Texas, 79409-1051, USA
}
\author{Ra Inta}
\altaffiliation[Current address: ]{
Accelebrate,
925B Peachtree Street, NE,
PMB 378,
Atlanta, GA 30309-3918,
USA
}
\affiliation{
Department of Physics and Astronomy,
Texas Tech University,
Lubbock, Texas, 79409-1051, USA
}

\begin{abstract}

We present the first searches for gravitational waves from $r$-modes of the Crab
pulsar, coherently and separately integrating data from three stretches of the
first two observing runs of Advanced LIGO using the $\mathcal{F}$-statistic.
The second run was divided in two by a glitch of the pulsar roughly halfway
through.  The frequencies and derivatives searched were based on radio
measurements of the pulsar's spin-down parameters as described in Caride
\textit{et al.,} Phys.\ Rev.\ D \textbf{100}, 064013 (2019).  We did not find
any evidence of gravitational waves. Our best 90\% confidence upper limits on
gravitational wave intrinsic strain were $1.5\times10^{-25}$ for the first
run, $1.3\times10^{-25}$ for the first stretch of the second run, and
$1.1\times10^{-25}$ for the second stretch of the second run. These are the
first upper limits on gravitational waves from $r$-modes of a known pulsar to
beat its spin-down limit, and they do so by more than an order of magnitude in
amplitude or two orders of magnitude in luminosity.

\end{abstract}

\maketitle

\acrodef{EM}{electromagnetic}
\acrodef{GW}{gravitational wave}
\acrodef{GWOSC}{Gravitational Wave Open Science Center}
\acrodef{LIGO}{Laser Interferometer Gravitational-wave Observatory}
\acrodef{O1}{first observing run}                                               
\acrodef{O2}{second observing run}
\acrodef{SFT}{short Fourier transform}

\section{Introduction}

\begin{table*} 
\begin{ruledtabular}
\begin{tabular}{lccccc} 
\textrm{Search} & \textrm{Start time (UTC)} & \textrm{End time (UTC)} &
$T_\mathrm{span}$ (days) & $T_\mathrm{dat}$ (days) & \textrm{SFTs}\\ 
\colrule
O1 & 09/12/15 06:03:57 & 01/19/16 15:34:47 & 129.4 & 133.1 & 6389
\\ 
O2 (early) & 11/30/16 18:01:57 & 03/27/17 16:28:25 & 117.1 &
128.9 & 6186 \\ 
O2 (late) & 03/28/17 23:47:38 & 08/25/17 21:59:34 & 150.9 &
167.4 & 8035 \\ 
\end{tabular} 
\end{ruledtabular}
\caption{Start and end times of the three searches.
The time difference between start and end is $T_\mathrm{span},$ while
$T_\mathrm{dat}$ is the live time of the data.
The latter can be greater than the former because there are two
interferometers.}
\label{table:Time} 
\end{table*}

Rapidly rotating neutron stars might be detectable emitters of long lived
quasi-monochromatic radiation known as continuous
\acp{GW}~\cite{Glampedakis2018}. The emission mechanism for continuous waves
could be a nonaxisymmetric mass quadrupole (``mountain'')
or a current-quadrupolar $r$-mode.
Hence the detection of continuous \acp{GW} might help reveal the underlying
properties of neutron star interiors. The \ac{GW} frequencies of many pulsars
lie in the most sensitive band of the \ac{LIGO}, so these rapidly rotating
neutron stars are attractive targets for continuous \ac{GW}
searches~\cite{Riles2017}.

The $r$-modes, whose frequencies are mainly determined by the Coriolis force,
are unstable to \ac{GW} emission~\cite{Andersson_1998,Friedman_1998} even
allowing for various damping mechanisms~\cite{LMoO:1998prl}.
Hence they might amplify and sustain
themselves, and might be the most interesting possibility for continuous
\acp{GW}. $R$-modes might play an important role in the spin-downs of the
fastest young neutron stars~\cite{Owen_1998} and in the regulation of spin
periods of some older accreting neutron stars~\cite{Bildsten1998,
Andersson:1998qs}.
Comparison of the \ac{GW}
frequency to the spin frequency determined from timing radio or x-ray pulses
could measure the compactness of a pulsar~\cite{Idrisy:2014qca}, and the
existence of $r$-modes at certain frequencies could constrain the properties of
the neutron superfluid~\cite{Kantor2020}.

Some \ac{GW} searches, starting with Ref.~\cite{Abadie_2010}, have set upper
limits on $r$-mode \ac{GW} emission. However most of these have been broad band
searches for neutron stars not previously known as pulsars. The searches
themselves did not take any extra steps to account for $r$-mode rather than
mass-quadrupole emission; rather the results could be interpreted in terms of
$r$-modes~\cite{Owen_2010}.

Caride \textit{et al.}~\cite{Caride2019} showed that dedicated searches for
$r$-modes from known pulsars are feasible. The key is to search the right range
of frequencies and frequency derivatives, which are significantly different from
the more often considered case of mass-quadrupole emission. Not surprisingly,
as with other \ac{GW} searches for pulsars, the Crab is the first prospect to
beat the spin-down limit using \ac{LIGO} data. The spin-down limit assumes that
all the rotational energy is lost in the form of \acp{GW}, and represents a
milestone a search must beat to have a chance of detection. Recently Fesik and
Papa~\cite{Fesik2020} first published an $r$-mode pulsar search similar to that
proposed by Caride \textit{et al.}~\cite{Caride2019}, for another pulsar which
is interesting for different reasons; but they did not beat its spin-down limit.
The Crab is a relatively nearby pulsar with one of the fastest known spin-down
rates, and thus it has one of the highest spin-down limits. Its rotational
frequency changes at the rate
$-3.69\times10^{-10}$\,Hz/s~\cite{JodrellBankObservatory, Lyne1993}. The Crab's pulse
timing is constantly measured by \ac{EM} observations, so the spin frequency
evolution of the Crab is well known during the LIGO \ac{O1} and \ac{O2}.

For various reasons we know that the Crab is not emitting \ac{GW} at or near the
spin-down limit.
\citet{Alford_2014} have argued that, if the $r$-mode instability operates
similarly in all neutron stars, the Crab is probably spinning to slowly to be
unstable in the presence of common damping mechanisms.
Observations of the Crab nebula indicate that most of the
rotational energy is lost powering the nebula via synchrotron and inverse
Compton radiation from the pulsar wind, and a few percent is lost in the narrow
light beam~\cite{B_hler_2014}. Recent \ac{GW} searches~\cite{Abbott_2019, O3}
have concluded that less than 0.01\% of the Crab's rotational energy loss is
through mass-quadrupole gravitational radiation. The braking index $n$ of the
Crab (the logarithmic derivative of its spin-down with respect to frequency) is
2.519.  This is closer to the $n=3$ expected for magnetic dipole
radiation~\cite{Lyne_2014} than to the $n=7$ expected for $r$-mode
emission~\cite{Owen_1998}. Such a low braking index is another indicator that
\ac{GW} emission is a small fraction of the spin-down limit. How small has not
been quantified in a model-independent way or for $r$-modes. But
Palomba~\cite{Palomba2000} found that, for a class of reasonable mass-quadrupole
models, the measured braking index of the Crab means it is emitting \ac{GW} at
least a factor of a few below the spin-down limit. We take this to suggest that
any $r$-mode \ac{GW} signal from the Crab must be at least a factor of a few (in
strain) below the spin-down limit.
Therefore a \ac{GW} search must beat the spin-down limit by a factor of a few in
amplitude (an order of magnitude in luminosity) to be interesting.

Nevertheless, Caride \textit{et al.}~\cite{Caride2019} showed that a search of
the Crab with interesting sensitivity is feasible, and we confirm this.
We performed searches for the Crab in \ac{O1} and \ac{O2} data (the publicly
available LIGO data sets at the time of writing) using the matched
filtering-based technique known as the $\mathcal{F}$-statistic.
While we did not find any evidence of a \ac{GW} signal, we were able to set
upper limits beating the spin-down limit by an interesting amount over a wide
parameter space.

\section{R-mode search method} 

Our search is based on a matched filtering technique  known as the 
$\mathcal{F}$-statistic. Developed by ~\citet{Jaranowski_1998} for a single interferometer and by
~\citet{PhysRevD.72.063006} for multiple interferometers, it is a statistical procedure
for the detection of the continuous gravitational waves. The
$\mathcal{F}$-statistic accounts for the amplitude modulation due
to the daily rotation of Earth in a computationally efficient
manner. In the presence of a signal,
$2\mathcal{F}$ is a non-central chi squared distribution with four degrees
of freedom and the non-centrality parameter is approximately the power
signal-to-noise ratio.
The
$\mathcal{F}$-statistic is one half the log of the likelihood function maximized over the
unknown strain, phase constant, inclination angle and polarization angle. 
The main issue when using the $\mathcal{F}$-statistic for this type of search,
as described by Caride \textit{et al.}~\cite{Caride2019}, is to
find the ranges of frequencies and frequency derivatives to search.

Pulsars are slowly spinning down due to \ac{GW} emission and other losses of rotational
energy. The evolution of the rotational frequency $\nu$ of a spinning down neutron star in the
frame of the solar system barycenter is approximated by
\begin{equation}
 \nu(t) = \nu\left( t_0 \right) + \dot{\nu}\left( t_0 \right) \left( t-t_0
\right) + \frac{1}{2} \ddot{\nu}\left( t_0 \right) \left( t-t_0 \right)^2,
\end{equation} 
where $t_0$ is a reference time (often the start time of the observation) and
dots indicate time derivatives.
Generally $\dddot{\nu}$ is not needed for less than a year of integration
time~\cite{Caride2019}. The frequency evolution is precisely known from
electromagnetic observations. It might depart from the above approximation due
to timing noise or glitches. Glitches are sudden increases in spin frequency
followed by exponential recovery to the pre-glitch
frequency~\cite{Espinoza_2011}.  Since the Crab glitched during O2, we divided
that search into pre- and post-glitch stretches. More on the Crab pulsar timing for
our searches will be discussed in Section~\ref{timing}.  Timing noise is residual
phase wandering of pulses relative to the normal spin down model.  Timing noise
will deviate \ac{GW} phases from Taylor series for time scales of a year or
longer~\cite{Jones_2004}. Assuming the \ac{GW} timing noise is similar to the
one observed in EM pulses, the mismatch of the Crab ephemeris during LIGO S5 run
from the model with no timing noise is less than
$1\%$~\cite{PhysRevD.91.062009}.  So, for all our searches (about four months of
data each), timing noise should not significantly mismatch the templates from
signal.

For a Newtonian star with spin frequency $\nu$, the $r$-mode \ac{GW} frequency
is approximately $f=\frac{4}{3}\nu$~\cite{Papaloizou1978}.
The frequency ratio deviates from $4/3$
when corrections due to general relativity, rapid rotation, superfluidity,
magnetic fields and core-crust coupling are considered~\cite{Idrisy:2014qca}.
For fast rotating neutron stars, the elastic restoring force on the crust
couples with the Coriolis restoring $r$-modes resulting in avoided
crossings~\cite{Levin_2001}. These are small frequency bands where the simple
relation between f and $\nu$ is drastically altered as modes change identities.
Away from avoided crossings, the $r$-mode frequency as a function of spin
frequency is approximately
\begin{equation}
f=A\nu-B\left(\frac{\nu}{\nu_K}\right)^2\nu, 
\end{equation} 
where $\nu_K$ is the Kepler frequency.
Here, as in Ref.~\cite{Caride2019}, we neglect effects other than slow rotation
and general relativity.  The effect of
rotation is to decrease the mode frequency in an inertial frame, so $A$ and $B$
are positive.

The ranges of $A$ and $B$ are chosen as in Ref.~\cite{Caride2019}:
We consider a range of compactness of
neutron stars ($0.11\leq M/R\leq 0.31$)~\cite{Idrisy:2014qca} which comes from
the uncertainty in the equation of state. The range of compactness gives a range
$1.39\leq A \leq 1.57$. The range of $B$ is derived from the relation of $f/\nu$
to the ratio of the rotational energy ($T$) to the gravitational potential energy
($W$)~\cite{Yoshida:2004gk}. For different polytropic indices and
compactnesses,
the range is 1.23--1.95 times the rotational parameter ($T/W$). \citet{Caride2019}
converted $T/W$ into  $(\nu/\nu_k)^2$ using  $M/R=0.1$ which gives a maximum
value of $B$ = 0.195. The minimum value of $B$ is not well known, so it is taken
to be zero. 

Our search covers the parameters ($f$, $\dot{f}$, $\ddot{f}$). 
The parameter ranges for our searches are given by \citet{Caride2019} based on
the considerations above:
\begin{eqnarray}
\label{frange}
\nu \left[ A_{\min} - B_{\max} \left( \nu/\nu_K \right)^2 \right] & \le & f
\le \nu\, A_{\max},
\\
\dot\nu \left[ f/\nu - 2B_{\max} \left( \nu/\nu_K \right)^2 \right] & \le &
\dot f \le \dot\nu f/\nu,
\\
0 & \le & \ddot f \le \ddot\nu f/\nu,
\end{eqnarray}
where, $f$ is spin frequency, $\dot{f}$ and $\ddot{f}$ are the first and second
spin derivative, $A_{min}=1.39,$ $A_{max}=1.57,$ and $B_{max}=0.195.$

We use the parameter space metric $g_{ij}$ to control the computational
cost of our search. 
The proper distance between two templates is given by the
mismatch ($\mu$), which is the loss in signal-to-noise ratio when signal falls
exactly between two template waveforms~\cite{Owen:1995tm}.  
Templates used to filter the data are chosen with a spacing determined by this
mismatch.
The template number is given by dividing the proper volume of the parameter
space by the proper volume per template~\cite{Caride2019}:
\begin{equation}\label{templatescalc}
\frac{\sqrt{g} \nu \left| \dot\nu \right| \ddot\nu B_{\max} \left(\nu/\nu_K
\right)^2 \left[ A_{\max}^2 - A_{\min}^2 \right]}{\left( 2 \sqrt{\mu/3}
\right)^3}
\end{equation}
where $g$ is the (constant) metric determinant.
The parameter space metric is given by~\cite{Wette:2008hg,Owen:1995tm}
\begin{equation}
g_{ij}=\frac{4\pi^2T_\mathrm{span}^{i+j+2}(i+1)(j+1)}{(i+2)!(j+2)!(i+j+3)},
\end{equation}\\
where $i$, $j$ stand for parameters $(f,\dot{f},\ddot{f})=(0,1,2)$ and
$T_\mathrm{span}$ is the time spanned by the observation. 

We look at the highest values of the $\mathcal{F}$-statistic
($2\mathcal{F}^*$) that survived the automated vetoes (described below).
The probability that a
given value of $2\mathcal{F}^*$ will be observed when no signal is present is
given by~\cite{Abadie_2010}: 
\begin{equation} 
P(N;2\mathcal{F}^*)= NP(\chi^2_4;2\mathcal{F}^*)[P(\chi^2_4;2\mathcal{F}^*)]^{N-1} 
\end{equation} 
where $N$ is the number of templates (assuming statistically independence) and
$P(\chi^2_4;2\mathcal{F}^*)$ is the central $\chi^2$ cumulative distribution
function with four degrees of freedom.
The above false alarm probability and corresponding threshold $\mathcal{F}^*$
assume gaussian noise, which is not always the case.
A high
$2\mathcal{F}$ is not enough to claim the detection, as instrumental lines might
act as a periodic signal. We follow each search with the
interferometer consistency veto as in (e.g.) Ref.~\cite{Lindblom_2020}, which
discards candidates when the joint $2\mathcal{F}$
value from the two interferometers is less than for either single interferometer.

\section{The Crab pulsar} \label{timing}

\begin{table} 
\begin{ruledtabular}
\begin{tabular}{lccc}
\textrm{Search} & \textrm{$\nu$} (Hz) &\textrm{$\dot{\nu}$} (Hz/s) &
\textrm{$\ddot{\nu}$} (Hz/s$^2$) \\ 
\colrule
O1 & $29.66181$ & $-3.693830\times10^{-10}$ &  $2.41\times10^{-20}$\\ 
O2 (early) & $29.64761$ & $-3.689673\times10^{-10}$ & $1.92\times10^{-20}$\\ 
O2 (late) & $29.64380$ & $-3.688438\times10^{-10}$ &  $2.5\times10^{-20}$\\ 
\end{tabular} 
\end{ruledtabular}
\caption{Timing of the Crab pulsar at the beginning of our three different searches.
The timing is measured by Jodrell Bank Observatory~\cite{Lyne1993}
and interpolated to the start time of each LIGO run. The displayed $\ddot{\nu}$ is the
maximum monthly value observed during each run.}
\label{table:timing} 
\end{table} 

The Crab pulsar is the remnant of a supernova explosion seen by
Chinese astronomers in the year 1054 AD. This is our first target due to its
high spin down limit, which is well above the LIGO \ac{O1} sensitivity
curve~\cite{Caride2019}. The rotational energy loss of the
pulsar is given by $\dot{E}=4\pi^2I_{zz}\nu\dot{\nu}\approx
4.33\times10^{31}$\,W, 
where ($I_{zz}$ = $10^{38}$\,kg\,m$^2$) is the principal moment of inertia ~\cite{Abbott_2008}. 
The spin-down power of the Crab is high enough that, even if
$r$-mode gravitational wave emission is only responsible for a small fraction of
it, the \acp{GW} could be detectable~\cite{Caride2019}.
 
The sky location of the Crab pulsar in J2000 coordinates is~\cite{Lyne1993} 
\begin{equation}
\begin{aligned} 
\alpha =05^h34^m31.94,^s\\ 
\delta=
+22^\circ00^\prime52.1.^{\prime\prime} 
\end{aligned} 
\end{equation}
 
The spin frequency and its derivative~\cite{Lyne1993} were obtained
from the Jodrell Bank observatory monthly ephemeris and interpolated to the
start dates of the LIGO run from the nearest dates of 09/16/2015 (\ac{O1}),
11/23/2016 (early O2) and 04/16/2017 (late O2). There were no glitches of the
Crab during \ac{O1}, so the whole run was coherently integrated. The Crab
glitched at 2017-03-27 22:04:48.000 UTC~\cite{Espinoza_2011} during the \ac{O2}
run, approximately halfway through. So we divided the \ac{O2} run into two
roughly equal stretches.  The glitch was very small by Crab standards
($\frac{\Delta\nu}{\nu}= 2.14\times 10^{-9}$)~\cite{Espinoza_2011},
so we started the search of late \ac{O2} a few hours afterwards although the
post-glitch relaxation time could be days.
Our code does not yet have the ability to combine pre- and post-glitch
stretches, so we kept the searches separate and present them as such.

Fast spinning down pulsars are contaminated by timing
noise~\cite{Lyne1992GlitchesAP}, especially when $\ddot{\nu}$ appears negative.
During
our observational time, the monthly $\ddot{\nu}$ measurement was fluctuating.
This might be due to external torque from the magnetosphere and might not affect
the high density interior which sources \acp{GW}. So we chose the
maximum $\ddot{\nu}$ of our search to be the maximum value observed during the observational time and
the minimum $\ddot{\nu}$ to be zero. The extra number of templates required by
including $\ddot{\nu}$ is factor of a few. Thus the searches were not too
expensive due to our probably over-wide range of $\ddot{\nu}.$

\section{Search Implementation} 

\begin{figure}
\includegraphics[width=3in]{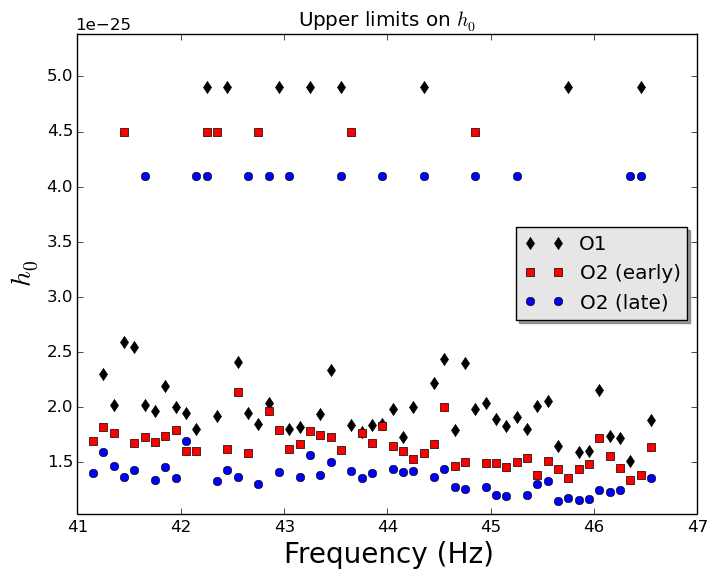}
\caption{
Upper limits (90\% confidence) on intrinsic strain vs.\ frequency, in 0.1\,Hz
bands, for our three searches are plotted below the legend.
The points arranged in straight lines above the legend indicate ``bad'' bands
(see text).
}
\label{fig:UL}
\end{figure}

We used data from the \ac{GWOSC}~\cite{collaboration2019open}, starting with time-domain strain data sampled
at 4\,kHz.  We downloaded all such \ac{O1} data for the official duration of the
run (from GPS times 1126051217 to 1137254417) from both interferometers (H1 and
L1), gating on only ``CBC CAT1'' vetoes. These indicate disastrous conditions
for the instruments, such as loss of laser power. We ignored the other vetoes
used in searches for binary black holes and neutron stars because they are aimed
at short-duration disturbances which are not significant for continuous wave
searches.  We then used the code \texttt{lalapps\_MakeSFTDAG} from
LALSuite~\cite{LALSuite} to
generate 1800\,s long \acp{SFT}, version\,2 format, high pass filtered with a knee
frequency of 7\,Hz and windowed with the default Tukey window. This produced
6,389 \acp{SFT} for \ac{O1} (3,474 from H1 and 2,915 from L1). A similar
procedure for \ac{O2} produced 14,231 \acp{SFT} (7,242 from H1 and 6,989 from
L1). Our $r$-mode searches used a modified version of a code usually used to
search for supernova remnants, most recently in Ref.~\cite{Lindblom_2020},
and was performed on the Texas Tech University
High Performance Computing Center's ``Quanah'' cluster.

The frequency bands of the searches were 41.2--46.6\,Hz (O1) and 41.1--46.6\,Hz (O2).
The minimum and maximum frequencies were rounded down and up respectively
from the range in Eq.~(\ref{frange})
because we used upper limit bands of a uniform  0.1\,Hz. We used a
bank of templates with minimal match $\mu=0.2.$ The ideal
template numbers for O1, early O2 and late O2 are $2.3\times 10^9$,
$9.7\times 10^8$ and $4.0\times 10^9$ respectively. The actual search produced
$1.7\times 10^{10}$, $1.2\times 10^{10}$ and $3.6 \times 10^{10}$ templates.
The number of templates of each search is around an order of magnitude greater than
the ideal number. This is because the code uses extra templates
to cover the boundaries outside the parameter space~\cite{Abadie_2010}. The
O1, early O2 and late O2 searches used computational times of 1584, 944 and 3892 core
hours on Quanah respectively.

Unlike in previous work such as Ref.~\cite{Lindblom_2020}, we did not use the
``Fscan veto,'' based on spectrograms of the data, to eliminate candidates
caused by nonstationarity and spectral lines.
Applying that veto resulted in the elimination of almost 1/4 of the frequency
band of each search.
This was due to the wide ``wings'' of the veto accounting for varying Doppler
shifts, spin-down values, and the width of the Dirichlet kernel (used in
computing $2\mathcal{F}$).
However, the same factors mean that template waveforms have less overlap with
stationary instrumental lines than in shorter searches done previously, and so eliminating
this veto did not produce an unmanageable number of candidate signals (see
below).

After using the interferometer consistency veto, we considered as candidates
batch jobs which produced $2\mathcal{F}$ values exceeding a threshold corresponding to a 5\%
false alarm probability (95\% detection confidence) in gaussian noise.
We found 29 search jobs with candidates:
16~in \ac{O1}, 7~in early \ac{O2}, and 6~in late \ac{O2}.
For each such job we inspected $2\mathcal{F}$ histograms and plots of
$2\mathcal{F}$ vs.\ frequency as in Ref.~\cite{Aasi_2015} and later works based
on it.
All candidate jobs showed distorted histograms and wide band disturbances,
indicating instrumental rather than astrophysical origin.
Many events just barely passed the interferometer consistency veto, which is a
lenient veto.
Although we did not use the known spectral lines (and combs) described in
Ref.~\cite{Covas2018} as \textit{a priori} vetoes, we found that many of our
candidates were coincident with those lines or with new lines in the combs
extending beyond those listed in Ref.~\cite{Covas2018}.

Therefore we do not claim any detection of gravitational waves.

\section{Upper limits}

\begin{figure}
\includegraphics[width=3in]{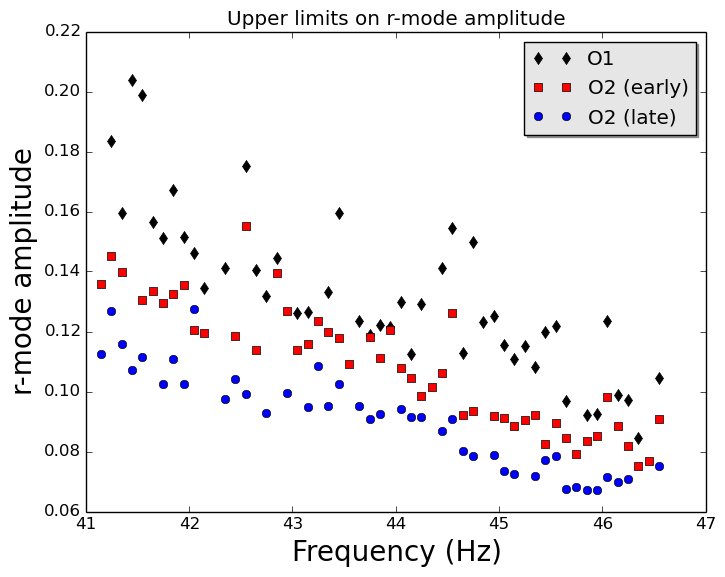}
\caption{
Upper limits (90\% confidence) on $r$-mode amplitude vs.\ frequency, in 0.1\,Hz
bands, for our three searches.
``Bad'' bands (see text) have been dropped.
}
\label{fig:alpha}
\end{figure}

In the absence of a detection, we set upper limits on gravitational wave
emission. The method is similar to Ref.~\citet{Lindblom_2020} among others.
Upper limits are the
weakest signal that can be detected from our search with a certain probability,
in our case chosen to be 90\%.
This means we set the false dismissal rate to $10\%$, and the loudest
$2\mathcal{F}$ observed (even if vetoed) will set the false alarm
rate~\cite{Aasi_2015}. Upper limit frequency bands are chosen to be small enough
for the interferometer noise to stay reasonably constant. The upper limit band
for all the searches is 0.1\,Hz. 

First, we use a computationally inexpensive Monte Carlo integration to find the
intrinsic strain amplitude ($h_0$) that exceeds the loudest observed
$2\mathcal{F}$ 90\% of the time.
This includes marginalization over inclination and polarization angles, which
reduce the actual strain amplitude in the data compared to $h_0.$
Consistent with the $\mathcal{F}$-statistic (and with most other directed
searches), we used priors corresponding to an isotropic probability distribution
of the pulsar's rotation axis.
In the case of the Crab the axis is known fairly well~\cite{Ng2008} and this
information could be used to improve the sensitivity of a \ac{GW}
search~\cite{Jaranowski2010}.
However the LALSuite code does not have this capability.

After the Monte Carlo, we use computationally expensive
($20$--$30\%$ of the cost of search)~\cite{Aasi_2015} software injection searches
to validate the upper limit. 
In each upper limit band we inject 1000 signals with various $h_0.$
For each $h_0$ we use variable $f,\dot{f},\ddot{f}$ and
inclination and polarization angles, and consider an injection ``detected'' if
it produces $\mathcal{F} > \mathcal{F}^*.$
Comparing this to the Monte Carlo is another check for contaminated frequency
bands.

Our upper limits on intrinsic strain in 0.1\,Hz bands for our three searches
are shown in Fig.~\ref{fig:UL}.
In some bands injections indicated that the true false dismissal rate was higher
than 10\%, typically due to many and/or strong spectral lines.
In Fig.~\ref{fig:UL}, for the sake of visibility, these bad bands are given
constant values well above the upper limits and thus appear along horizontal
lines above the figure legend.
The data files are included in the supplemental material to this
article~\cite{supp}.
The best $90\%$ confidence level upper limits on
intrinsic strain amplitude are $1.5\times 10^{-25}$ (\ac{O1}),
$1.3\times 10^{-25}$ (early \ac{O2}) and $1.1\times 10^{-25}$ (late \ac{O2}).
\ac{O2} upper limits beat the spin-down limit by more than an order of
magnitude.
Late \ac{O2} was the most sensitive because it had the most data and the noise
spectrum was better (lower) than the other searches.
Upper limits can also be characterized in terms of a statistical factor
$\Theta$~\cite{Wette:2008hg} of the form
\begin{equation}
h_0 = \Theta \sqrt{S_h / T_\mathrm{dat}},
\end{equation}
where $T_\mathrm{dat}$ is the data live time and $S_h$ is the
power spectral density of strain noise.
We achieved an average $\Theta \simeq 36,$ typical for coherent directed
searches.
Another figure of merit is the ``sensitivity depth''~\cite{Dreissigacker2018}
\begin{equation}
\mathcal{D} = \sqrt{S_h}/h_0 = \sqrt{T_\mathrm{dat}} / \Theta.
\end{equation}
Our searches achieve $\mathcal{D} \simeq 100$\,Hz$^{-1/2},$ comparable to narrow
band searches for known pulsars (among the best directed searches).

We can also set upper limits on $r$-mode amplitude.  
The coversion of $h_0$ to $r$-mode amplitude ($\alpha$) is shown by \citet{Owen_2010},
who took the fiducial value of moment of inertia ($I_{zz}$ = $10^{38}$\,kg\,m$^2$)
and typical $M=1.4\,M_\odot$. We convert the upper limit on
intrinsic strain $h_0$ to $r$-mode amplitude $\alpha$ using
\begin{equation} 
\alpha
= 0.028 \left( \frac{h_0} {10^{-24}} \right) \left( \frac{r} {\mbox{1 kpc}}
\right) \left( \frac{\mbox{100 Hz}} {f} \right)^3.
\end{equation} 
Upper limits on $\alpha$ are plotted (without the bad bands) in
Fig.~\ref{fig:alpha}.
The best $r$-mode amplitude upper limits for our different searches are $0.085$
(O1), $0.075$ (early O2) and $0.067$ (late O2). 

\section{Conclusion} 

Our searches of \ac{LIGO} \ac{O1} and \ac{O2} data for the Crab pulsar did not
detect $r$-mode \acp{GW}.
However, we set the first upper limits on $r$-mode \ac{GW} emission from this
pulsar.
These upper limits beat the spin-down limit by an order of magnitude in strain
or two orders of magnitude in luminosity, and are the first to do so for
$r$-modes from a known pulsar.
The corresponding upper limits on $r$-mode amplitude are not competitive with
most predictions of $r$-mode saturation in young neutron stars such as
Ref.~\cite{Bondarescu2009}.

In the near future, with more and better data, we will be able to extend this
type of search to longer data sets with lower noise, thereby increasing
sensitivity.
We will also be able to beat the spin-down limits of more pulsars.
With improvements in our code, we will be able to take advantage of the spin
axis orientation for those pulsars (such as the Crab) for which it is known and
integrate data sets with glitches in them.

\acknowledgments

We are grateful to various members of the LSC continuous waves working group,
especially Ian Jones and Karl Wette, for helpful discussions over the years.
This work was supported by NSF grant PHY-1912625.
This research has made use of data, software and/or web tools obtained from the
Gravitational Wave Open Science Center (\url{https://www.gw-openscience.org}), a
service of LIGO Laboratory, the LIGO Scientific Collaboration and the Virgo
Collaboration. LIGO is funded by the U.S. National Science Foundation. Virgo is
funded by the French Centre National de Recherche Scientifique (CNRS), the
Italian Istituto Nazionale della Fisica Nucleare (INFN) and the Dutch Nikhef,
with contributions by Polish and Hungarian institutes. The authors acknowledge
the High Performance Computing Center (HPCC) at Texas Tech University for
providing computational resources that have contributed to the research results
reported within this paper (\url{http://www.depts.ttu.edu/hpcc/}).
An earlier draft of this material was included in the first author's Ph.D.\
thesis (Texas Tech University, 2020, unpublished).

\bibliography{rmodesrch} 

\begin{thebibliography}{44}
\bibitem [{sup()}]{supp}%
  \BibitemOpen
  \href@noop {} {}\bibinfo {note} {PRD will insert URL here.}\BibitemShut
  {Stop}%
\end{thebibliography}
\end{document}